\documentclass{article}

\begin{document}

\noindent \textbf{The manifestly covariant Aharonov-Bohm effect in terms of
the 4D fields\bigskip }$\bigskip $

\noindent Tomislav Ivezi\'{c}$\bigskip $

\noindent \textit{Ru%
\mbox
 {\it{d}\hspace{-.15em}\rule[1.25ex]{.2em}{.04ex}\hspace{-.05em}}er Bo\v{s}%
kovi\'{c} Institute, P.O.B. 180, 10002 Zagreb, Croatia}

\noindent E-mail: \textit{ivezic@irb.hr}\bigskip \medskip

\noindent In this paper it is presented a manifestly covariant formulation
of the Aharonov-Bohm (AB) phase difference for the magnetic AB effect . This
covariant AB phase is written in terms of the Faraday 2-form $F$\ and using
the decomposition of $F$\ in terms of the electric and magnetic fields as
four-dimensional (4D) geometric quantities. It is shown that there is a
static electric field outside a stationary solenoid with resistive conductor
carrying steady current, which causes that the AB phase difference in the
magnetic AB effect may be determined by the electric part of the covariant
expression, i.e., by the local influence of the 4D electric field and not,
as generally accepted,\emph{\ }in terms of nonzero vector potential.\bigskip
\medskip

\bigskip \bigskip

\noindent \textbf{1. Introduction}\bigskip

In a recent paper [1] the covariant generalizations of the Aharonov-Bohm
(AB) effect [2] are considered. One of these generalizations, which will be
investigated in this paper, is in terms of the space-time \textquotedblleft
area\textquotedblright\ integral of the electric and magnetic fields written
in terms of the Faraday 2-form $F$, Eq. (6) in [1] or Eq. (\ref{cf}) here.

In this paper two important changes relative to [1] will be presented. The
first change, which will be discussed in Sec. 2, refers to the mathematical
formulation, whereas the second one refers to the physical interpretation of
the AB phase shift and it will be discussed in Sec. 3, see also Sec. 10 in
[3]. It is true that the expression for the AB phase difference, Eq. (6) in
[1], is a covariant expression, but it is not the case with the
decomposition of $F$ in terms of the components of the 3-vectors $\mathbf{E}$
and $\mathbf{B}$, Eq. (7) in [1]. Instead of it a manifestly covariant
decomposition of $F$, i.e., of $F_{\mu \nu }$, will be presented by Eq. (\ref%
{FE}). As can be seen from [4-9], in the four-dimensional (4D) spacetime, in
contrast to the usual transformations (UT) of the 3-vectors $\mathbf{E}$ and
$\mathbf{B}$, Eq. (\ref{ee}) here, or Eq. (11.148) in [10], according to
which the transformed $\mathbf{E}^{\prime }$ is expressed by the mixture of
the 3-vectors $\mathbf{E}$ and $\mathbf{B}$, the mathematically correct
Lorentz transformations (LT) always transform the 4D algebraic object
representing the electric field only to the electric field; there is no
mixing with the magnetic field, Eq. (\ref{el}) here or Eqs. (42) and (43) in
[3]. This is first shown by Minkowski in Sec. 11.6 in [11] and reinvented
and generalized in terms of 4D geometric quantities in [4-9]. A brief
discussion is given in [3]. Using such 4D electric and magnetic fields the
manifestly covariant expression for the AB phase difference is given by Eq. (%
\ref{ps}), which replaces Eqs. (8) and (9) from [1].

In Sec. 3, we use the results from [3], particularly it refers to the
discussion from Sec. 10 in [3]. There, it is mentioned that always there are
external electric fields from stationary resistive conductors carrying
constant currents, see, e.g., Sec. 4 in [12] and references therein. In
Secs. 7-7.2 in [3] it is shown that in the 4D geometric approach to special
relativity, the invariant special relativity (ISR), there is a static
electric field outside a moving and a \emph{stationary} solenoid with a
steady current not only for resistive conductors but also for
superconductors. Note that in the ISR an independent physical reality is
attributed to the 4D geometric quantities and not, as usual, to the 3D
quantities. Furthermore, in Sec. 8 in [3], it is discovered that there is
such static 4D electric field not only outside a \emph{moving} permanent
magnet, as generally accepted in physical literature, but outside a \emph{%
stationary} permanent magnet as well. As explained in [3] that result is
based on the paper [13] in which the generalized Uhlenbeck-Goudsmit
hypothesis is formulated, Eq. (9) in [13], i.e., Eq. (59) in [3]. The
mentioned results for the existence of the 4D external electric fields may
give the possibility to explain the experimentally observed fringe shift for
the magnetic AB effect even in Tonomura's experiments [14], Sec. 10 in [3]:
\textquotedblleft in terms of forces, which so far have been
overlooked.\textquotedblright\ Here, in Sec. 3, these results from [3] are
combined with the correct covariant formulation of the AB effect from Sec.
2, i.e., with Eqs. (\ref{de}) and (\ref{dae}) for $\delta \alpha _{E}$, to
explain the existence of the magnetic AB phase difference in terms of the
overlooked 4D electric force and not, as usual, in terms of the vector
potential.

The existence of the overlooked 4D external electric fields is one of the
reasons why we do not consider the covariant AB phase in terms of the
four-potentials, $\delta \alpha _{EB}=(e/\hbar )\oint A_{\mu }dx^{\mu }$,
Eq. (5) in [1]. Another reason is that in [15] an axiomatic formulation of
the electromagnetism is presented in which only the field equation for $F$
is postulated, Eq. (4) in [15], i.e., Eq. (20) in [3]. It is shown in [15]
that the electromagnetic field $F$ can be taken as \emph{the primary quantity%
} for the whole electromagnetism both in the theory and in experiments;\emph{%
\ }$F$ is a well-defined 4D \emph{measurable} quantity. It yields the
complete description of the electromagnetic field and there is no need to
introduce either the potentials (thus dispensing with the need for the gauge
conditions) or the field vectors. That formulation with the $F$ field is a
self-contained, complete and consistent formulation. The generalization of
Eq. (4) in [15] to a moving medium is presented in [16]. There, [16], the
field equations are written in terms of $F$\ and the generalized
magnetization-polarization bivector $\mathcal{M}$ and not, as usual, in
terms of $F$\ and the electromagnetic excitation tensor $\mathcal{H}$%
.\bigskip \bigskip

\noindent \textbf{2. Covariant expression for the AB phase shift}\bigskip

The covariant expression for the AB phase difference in terms of the Faraday
2-form $F$ is presented by Eq. (6) in [1], which is repeated here%
\begin{equation}
\delta \alpha _{EB}=(-e/2\hbar )\int F_{\mu \nu }dx^{\mu }\wedge dx^{\nu
}=(e/\hbar )\int F,  \label{cf}
\end{equation}%
where $F=(-1/2)F_{\mu \nu }dx^{\mu }\wedge dx^{\nu }$. (The notation is the
same as in [1]; $dx^{\mu }$ and $dx^{\nu }$ are differential four-vectors
and throughout the paper we set $c=1$.) In order to show that this covariant
expression (\ref{cf}) reduces to the usual expressions with the 3-vectors,
Eqs. (2) and (4) in [1], the Faraday 2-form $F$ is decomposed using the
components of the 3-vectors $\mathbf{E}$ and $\mathbf{B}$, Eq. (7) in [1].
It is worth mentioning that Eq. (7) in [1] is not mathematically correct;
the expression $(-1/2)F_{\mu \nu }dx^{\mu }\wedge dx^{\nu }$ (it will be
denoted as (F)) is covariant under the LT, but it is not the case with its
decomposition $(E_{x}dx+E_{y}dy+E_{z}dz)\wedge dt+B_{x}dy\wedge
dz+B_{y}dz\wedge dx+B_{z}dx\wedge dy$ (it will be denoted as (EB)). The
expression (EB) is obtained from that one with $F_{\mu \nu }$, (F), using
the usual identification of the components of $F_{\mu \nu }$\ with the
components of the 3-vectors $\mathbf{E}$ and $\mathbf{B}$, e.g., Eq.
(11.138) in [10], see also Eq. (3) and the comment on it in [3]. In all
traditional approaches it is supposed that the same identification holds in
a relatively moving inertial frame of reference, see Eq. (7) in [3]. This
means that it is considered that the components of $\mathbf{E}$ and $\mathbf{%
B}$ transform under the LT as the components of $F_{\mu \nu }$\ transform,
i.e., that the LT of the components of $\mathbf{E}$ and $\mathbf{B}$ (for
the boost in the $x$ direction) are
\begin{eqnarray}
E_{x}^{\prime } &=&E_{x},\ E_{y}^{\prime }=\gamma (E_{y}-\beta B_{z}),\
E_{z}^{\prime }=\gamma (E_{z}+\beta B_{y}),  \nonumber \\
B_{x}^{\prime } &=&B_{x},\ B_{y}^{\prime }=\gamma (B_{y}+\beta E_{z}),\
B_{z}^{\prime }=\gamma (B_{z}-\beta E_{y}),  \label{ee}
\end{eqnarray}%
see, e.g., Sec. 11.10 and Eq. (11.148) in [10], or the discussion and
equations (9) and (10) in [3]. The essential feature of the transformations (%
\ref{ee}) is that \emph{the transformed components} $E_{x,y,z}^{\prime }$\
\emph{are expressed by the mixture of the components of the 3-vectors} $%
\mathbf{E}$ \emph{and} $\mathbf{B}$, \emph{and similarly for} $\mathbf{B}%
^{\prime }$. The electric field $\mathbf{E}$ in one inertial frame is
\textquotedblleft seen\textquotedblright\ as slightly changed electric field
$\mathbf{E}^{\prime }$ and an \emph{induced magnetic field} $\mathbf{B}%
^{\prime }$ in a relatively moving inertial frame. From the time of
Einstein's fundamental paper [17], the transformations (\ref{ee}) are always
considered to be the relativistically correct LT (boosts) of $\mathbf{E}$
and $\mathbf{B}$, but we shall call them, as already said, the UT. As can be
seen from Secs. 3.1 and 3.2 in [3], the above mentioned identification is
synchronization dependent and it holds only if Einstein's synchronization
[17] is used. There, it is also shown that the mentioned identifications are
meaningless if only the Einstein synchronization is replaced by an
asymmetric synchronization, the \textquotedblleft radio\textquotedblright\
synchronization. That nonstandard synchronization is described in more
detail in [18], see also [13]. This is also mentioned below, see Eq. (\ref%
{ptr}) and the discussion with it. But, \emph{different synchronizations are
only different conventions and physics must not depend on conventions.}

Therefore, as first shown by Minkowski in Sec. 11.6 in [11] and
independently reinvented and generalized in terms of the 4D geometric
quantities in [4-9], $F_{\mu \nu }$\ can be decomposed in a covariant manner
\begin{eqnarray}
F_{\mu \nu } &=&(v_{\mu }E_{\nu }-v_{\nu }E_{\mu })+\varepsilon _{\mu \nu
\alpha \beta }v^{\alpha }B^{\beta },  \nonumber \\
E_{\mu } &=&F_{\nu \mu }v^{\nu },\quad B_{\mu }=(1/2)\varepsilon _{\mu \nu
\alpha \beta }F^{\nu \alpha }v^{\beta },  \label{fc}
\end{eqnarray}%
where $E_{\mu }$ and $B_{\mu }$ are the components of the 4D electric and
magnetic fields respectively, whereas $v_{\mu }$ are the components of the
4D velocity of a family of observers who measure electric and magnetic
fields, see also Sec. 5 in [3]. Since $F_{\mu \nu }$\ is antisymmetric it
holds that $E_{\mu }v^{\mu }=B_{\mu }v^{\mu }=0$, only three components of $%
E_{\mu }$ and $B_{\mu }$ are independent. In the 4D spacetime the
mathematically correct decomposition of $F$\ into 4D electric and magnetic
fields \emph{and} the 4-velocity of the observer, Eq. (\ref{fc}), is already
firmly theoretically founded and it is known to many physicists. The recent
example is in [19]; it is only the electric part (the magnetic part is zero
there). Similarly, in the component form as in (\ref{fc}), this
decomposition is presented, e.g., in [20] and in the basis-free form with
the abstract 4D quantities, e.g., in [21].

From the mathematical viewpoint it is trivially to see how, e.g., $E_{\mu }$%
\ from (\ref{fc}) is transformed under the LT; in the mathematically correct
LT the transformed components $E_{\mu }^{\prime }$ are not determined only
by $F_{\mu \nu }^{\prime }$,\ as in all usual approaches, e.g., Eqs.
(11.147) and (11.148) in [10], but also by $v^{\prime \mu }$. This is first
shown by Minkowski in Sec. 11.6 in [11]. Let $v$, $E$ and $B$ are $1\times 4$
matrices and $F$ is a $4\times 4$ matrix; their components are implicitly
determined in the standard basis. Minkowski first described how $v$ and $F$
separately transform under the LT $A$ (the matrix of the LT is denoted as $A$%
). The LT of the 4-velocity $v$ is $v^{\prime }=vA$ and the LT of the
field-strength tensor $F$\ is $F^{\prime }=A^{-1}FA$, then, as shown by
Minkowski, the mathematically correct LT of $E=vF$ is $E=vF\longrightarrow
E^{\prime }=(vA)(A^{-1}FA)=(vF)A=EA$. This means that under the LT \emph{both%
} quantities, the field-strength tensor $F$\ ($4\times 4$ matrix) \emph{and
the 4-velocity} $v$ ($1\times 4$ matrix)\ are transformed and their product
transforms as any $1\times 4$ matrix transforms. As already stated that
mathematically correct procedure is reinvented and generalized using the 4D
geometric quantities both in the tensor formalism and in the geometric
algebra formalism in [4-9]. Particularly, the comparison with Minkowski's
results, Sec. 11.6 in [11], is presented in [9]. The essential point is that
\emph{the 4D electric field }$E$ \emph{transforms by the LT again to the 4D
electric field }$E^{\prime }$\emph{; there is no mixing with the 4D magnetic
field} $B$, i.e., \emph{the components} $E_{\mu }$ \emph{transform by the LT
again to the components} $E_{\mu }^{\prime }$ \emph{of the same 4D electric
field and there is no mixing with }$B_{\mu }$,%
\begin{equation}
E_{0}^{\prime }=\gamma (E_{0}+\beta E_{1}),\ E_{1}^{\prime }=\gamma
(E_{1}+\beta E_{0}),\ E_{2,3}^{\prime }=E_{2,3},  \label{el}
\end{equation}%
for a boost along the $x^{1}$ axis. It is easily seen that the UT, Eq.
(11.148) in [10], i.e., Eq. (\ref{ee}) here, will be simply obtained in this
4D geometric approach if \emph{only} the components $F_{\mu \nu }$\ are
transformed \emph{but not} the components $v^{\mu }$. Such procedure
corresponds to the usual identifications of the components of $F_{\mu \nu }$%
\ with the components of the 3-vectors $\mathbf{E}$ and $\mathbf{B}$ in both
relatively moving inertial frames of reference. A short derivation of these
results can be seen in [7]. In this case there is no need to write the
transformations for the components $B_{\mu }$\ since they transform as in (%
\ref{el}). This means that it is proved in [4-9] that, contrary to the
generally accepted opinion, \emph{the UT of the 3-vectors }$\mathbf{E}$
\emph{and} $\mathbf{B}$, Eq. (\ref{ee}), \emph{are not the LT}, but that the
mathematically correct LT are given by Eq. (\ref{el}). For a brief review
see Sec. 5 in [3] or Sec. 3 in [22]. It is interesting that although Eq. (%
\ref{fc}) is known to many physicists, e.g., [20, 21], it is not noticed
that the mathematically correct LT of, e.g., $E_{\mu }=F_{\nu \mu }v^{\nu }$%
, necessarily require that both $F_{\nu \mu }$\ and $v^{\nu }$\ have to be
transformed and not only $F_{\nu \mu }$.\ In the 4D spacetime, from the
mathematical viewpoint, the 4D electric and magnetic fields are correctly
defined and they transform as any other 4-vector transforms, i.e., according
to Eq. (\ref{el}).

Hence, instead of Eq. (7) in [1] we have
\begin{eqnarray}
F &=&(-1/2)F_{\mu \nu }dx^{\mu }\wedge dx^{\nu }=  \nonumber \\
&&(-1/2)[(v_{\mu }E_{\nu }-v_{\nu }E_{\mu })+\varepsilon _{\mu \nu \alpha
\beta }v^{\alpha }B^{\beta }]dx^{\mu }\wedge dx^{\nu }.  \label{FE}
\end{eqnarray}%
In Eq. (\ref{FE}) both expressions for $F$\ are manifestly covariant under
the LT, which does not hold, as already stated, for Eq. (7) in [1]. In
contrast to the usual treatment from [1], $\delta \alpha _{EB}$\ that is
given by Eq. (\ref{ps}) below \emph{is the same for all relatively moving
inertial observers and for all coordinate bases used by them; the principle
of relativity is naturally satisfied.} This proves a mathematical and
relativistic correctness of this manifestly covariant approach.

For the reader's convenience and for easier comparison with [1] we have
written, e.g., Eq. (\ref{fc}), only with components, but as $F$\ is a 4D
geometric quantity, a 2-form ($F=(-1/2)F_{\mu \nu }dx^{\mu }\wedge dx^{\nu }$%
), so is, e.g., the electric field $E$, a 4D geometric quantity, an 1-form ($%
E=E_{\mu }dx^{\mu }$). Both, $F$\ and $E$\ in these relations are written in
a specific coordinate basis, the standard basis, with the Einstein
synchronization of distant clocks and Cartesian space coordinates. In [1],
as in all usual covariant approaches, the standard basis is exclusively
used, but, as pointed out above, different systems of coordinates are
allowed in an inertial frame and they are all equivalent in the description
of physical phenomena. Thus, for example, one can use the above mentioned
asymmetric synchronization, the \textquotedblleft radio\textquotedblright\
synchronization. The important difference relative to the usual formulation
with 3-vectors is that in the 4D spacetime a 4D geometric quantity is \emph{%
the same 4D quantity} for all inertial observers and for all coordinate
bases used by them, $E=E_{\mu }dx^{\mu }=E_{\mu }^{\prime }dx^{\prime \mu
}=E_{\mu ,r}dx^{\mu ,r}=...$ , where the primed quantities are the Lorentz
transforms of the unprimed ones and the quantities with the index
\textquotedblleft $r$\textquotedblright\ are in the coordinate basis with
the \textquotedblleft radio\textquotedblright\ synchronization. Observe that
in [18] and in the second and third papers in [23] the \textquotedblleft
radio\textquotedblright\ synchronization is used throughout the papers.
Moreover, in Eq. (4) in [18] it is presented the transformation matrix that
connects Einstein's system of coordinates with another system of coordinates
in the same reference frame. Also, Eq. (1) in [18], it is derived such form
of the LT, which is independent of the chosen system of coordinates,
including different synchronizations. Since in the ISR every 4D geometric
quantity is invariant under the LT\ the principle of relativity is
automatically satisfied and there is no need to postulate it outside the
mathematical formulation of the theory as in Einstein's formulation of SR,
[17].

For simplicity and for easier comparison with [1] we shall introduce the
inertial frame of \textquotedblleft fiducial\textquotedblright\ observers ($%
v^{\mu }=(1,0,0,0)$) with the standard basis (Einstein's synchronization) in
it, which will be called the \textquotedblleft f\textquotedblright -frame.
In that frame it holds that $E_{0}=B_{0}=0$ and only the spatial components
of $E_{\mu }$\ and $B_{\mu }$\ remain. From (\ref{fc}) it follows that these
components are
\begin{equation}
E_{i}=F_{0i}v^{0}=F_{0i},\quad B_{i}=(1/2)\varepsilon _{0ijk}F^{kj};
\label{ffr}
\end{equation}%
the same components as in, e.g., Eq. (11.138) in [10]. Observe that the
\textquotedblleft f\textquotedblright -frame is not any kind of a preferred
frame, because any inertial frame can be chosen to be that frame and it is
usually taken that the laboratory frame is the \textquotedblleft
f\textquotedblright -frame. However, in any other relatively moving inertial
frame, the $S^{\prime }$ frame, the \textquotedblleft
fiducial\textquotedblright\ observers are moving, and the components $v^{\mu
}$ transform as in (\ref{el}), $v^{\prime \mu }=(\gamma ,-\beta \gamma ,0,0)$%
. Hence, as already shown by Minkowski in Sec. 11.6 in [11], for the
transformations from the \textquotedblleft f\textquotedblright -frame, see
[7], $(E_{\mu })^{\prime }=[F_{\nu \mu }v^{\nu }]^{\prime
}=[F_{0i}v^{0}]^{\prime }=F_{\nu \mu }^{\prime }v^{\prime \nu }=E_{\mu
}^{\prime }$, and Eq. (\ref{el}) is obtained; \emph{the components} $E_{\mu
} $ \emph{transform by the LT again to the components} $E_{\mu }^{\prime }$.
Let us take in (\ref{FE}) that $E_{1}=E_{x}$, ... , $B_{1}=B_{x}$, ... , $%
\varepsilon _{0123}=1$, $dx^{0}=dt$, .... , $dx^{3}=dz$, then in the
\textquotedblleft f\textquotedblright -frame the second covariant expression
in (\ref{FE}) corresponds to the expression (EB) that is used in [1]. In a
relatively moving inertial frame $S^{\prime }$ the LT (\ref{el}) will give
that $E_{0}^{\prime }$ and $B_{0}^{\prime }$ will be different from zero and
these terms cannot exist in the approach from [1], which deals with the
expression (EB), i.e., with the fields as the 3-vectors.

In the usual formulation the physical meaning of 3-vectors $\mathbf{E}$ and $%
\mathbf{B}$ is determined by the the Lorentz force as a 3-vector $\mathbf{F=}%
q\mathbf{E}+q\mathbf{u}\times \mathbf{B}$\ and by Newton's second law $%
\mathbf{F}=d\mathbf{p}/dt$, $\mathbf{p=}m\gamma _{u}\mathbf{u}$.

However, in the 4D spacetime, the Lorentz force $K$ is not a 3-vector, but
it is a 4D geometric quantity. $K$ is the contraction of the electromagnetic
2-form $F$\ with particle's 4-velocity $u$ (it is defined to be the tangent
to its world line). The components of $K$ in the standard basis are $K_{\mu
}=qF_{\mu \nu }u^{\nu }$, where $u^{\mu }$ is the 4-velocity (components) of
a charge $q$, or with $E_{\mu }$ and $B_{\mu }$, using the decomposition of $%
F_{\mu \nu }$, (\ref{fc}), they become%
\begin{equation}
K_{\mu }=q[(v_{\mu }E_{\nu }-v_{\nu }E_{\mu })+\varepsilon _{\mu \nu \alpha
\beta }v^{\alpha }B^{\beta }]u^{\nu }.  \label{km}
\end{equation}%
\emph{In the 4D spacetime, the physical meaning of} $E_{\mu }$ \emph{and} $%
B_{\mu }$ \emph{is determined by the Lorentz force} $K_{\mu }$\ \emph{and by
the 4D expression for Newton's second law} $K_{\mu }=dp_{\mu }/d\tau $, $\
p_{\mu }=mu_{\mu }$, where $p_{\mu }$ is the proper momentum (components)
and $\tau $\ is the proper time. All components $E_{\mu }$ and $B_{\mu }$,
thus $E_{0}$ and $B_{0}$ as well, are equally well physical and measurable
quantities by means of the mentioned $K_{\mu }$\ and the equation of motion,
i.e., the 4D expression for Newton's second law. Obviously, regardless of
the fact that majority of physicists believe that only the 3-vectors $%
\mathbf{E}$ and $\mathbf{B}$ are physical and measurable quantities, \emph{%
in the 4D spacetime, the 4D geometric quantities are properly defined both
theoretically and} \emph{experimentally}. In view of this discussion it is
obvious that the question what physically are $E_{0}$ and $B_{0}$ is
equivalent to the question - what is the temporal component $x_{0}$ of the
position 4-vector. This is particularly visible if the Einstein
synchronization is replaced by the \textquotedblleft
radio\textquotedblright\ synchronization in which the space and time are not
separated, see Eq. (\ref{ptr}) below. Then, the usual 3-vector $\mathbf{%
\mathbf{r}}$, and similarly the 3-vectors $\mathbf{E}$ and $\mathbf{B}$, are
meaningless. This fundamental difference between the usual formulation with
the 3D quantities and the formulation with the 4D geometric quantities is
exposed in much more detail, e.g., in [24].

It is also shown in, e.g., [5, 6, 15] that \emph{the LT of the 4D} $E_{\mu }$
\emph{and} $B_{\mu }$, (\ref{el}), \emph{are in a true agreement
(independent of the chosen inertial reference frame and of the chosen system
of coordinates in it) with all experiments in the electromagnetism,} \emph{%
whereas it is not the case with the UT of the 3D} $\mathbf{E}$ \emph{and} $%
\mathbf{B}$, (\ref{ee}). Thus, for example, it is shown in [5] that the
conventional theory with the 3D $\mathbf{E}$ and $\mathbf{B}$ and their UT
yields different values for the motional emf $\varepsilon $ for relatively
moving inertial observers, $\varepsilon =UBl$ and $\varepsilon =\gamma UBl$,
whereas the approach with 4D geometric quantities and their LT yields always
the same value for $\varepsilon $, which is defined as a Lorentz scalar, $%
\varepsilon =\gamma UBl$. This result is very strong evidence that the usual
approach is not relativistically correct, i.e., it is not in agreement with
the principle of relativity. It is on the experimentalists to find the way
to precisely measure the emf $\varepsilon $ for the considered problem of a
conductor moving in a static magnetic field and to see that in the
laboratory frame $\varepsilon =\gamma UBl$ and not simply $\varepsilon =UBl$%
. That problem is of a considerable importance in practice. The similar
discussion with the same conclusions was presented for the Faraday disk in
[6]. In the already mentioned [15] and in [25] the Trouton-Noble paradox is
considered. It is shown that in the geometric approach with 4D quantities
\emph{the 4D torques} will not appear for the moving capacitor if they do
not exist for the stationary capacitor, which means that with 4D geometric
quantities the principle of relativity is naturally satisfied and \emph{%
there is not the Trouton-Noble paradox}. The same conclusion holds in the
low-velocity approximation $\beta \ll 1$, or $\gamma \simeq 1$. It is also
shown in the same geometric approach with 4D torques that there is no
Jackson's paradox [24] and the \textquotedblleft charge-magnet
paradox\textquotedblright\ [22].

At this point it is worth noting that in the mathematically correct
approach, in general, there is no room for the 3-vectors in the 4D
spacetime. Let us better explain that statement. It is written in [1] after
Eq. (7) that: \textquotedblleft $F=B_{x}dy\wedge dz+B_{y}dz\wedge
dx+B_{z}dx\wedge dy=\mathbf{B}\cdot d\mathbf{S}$ where the differential
forms expression has been converted back to three-vector notation and $d%
\mathbf{S}$\ is the differential area.\textquotedblright\ However, such an
equality is mathematically impossible and incorrect. Namely, in the
mathematically correct formulation $dx$,$\ dy$, $dz$ have to be understood
as differential 4-vectors $dx^{1}$,$\ dx^{2}$,$\ dx^{3}$, respectively, the
4D geometric quantities that are properly defined on the 4D spacetime; the
wedge product refers to such 4D quantities and not to the usual scalar
differentials. On the other hand, $\mathbf{B}$ and $d\mathbf{S}$, as \emph{%
geometric quantities in the 3D space}, are constructed from the components
and \emph{the unit 3-vectors }$\mathbf{i}$, $\mathbf{j}$, $\mathbf{k}$,
e.g., $\mathbf{B=}B_{x}\mathbf{i}+B_{y}\mathbf{j}+B_{z}\mathbf{k}$. The unit
3-vectors have nothing to do with the basis in the 4D spacetime. The LT are
properly defined on the 4D spacetime and they cannot transform the
3-vectors. Hence, in the 4D spacetime it is not mathematically correct to
state as in [1]: \textquotedblleft .. the expression in (6) reduces to $%
\delta \alpha _{EB}=(e/\hbar )\int F=(e/\hbar )\int \mathbf{B}\cdot d\mathbf{%
S}$ which is equivalent to the 3-vector expression (2).\textquotedblright\
In the 4D spacetime the covariant expression ($(e/\hbar )\int F$) is the
correct one, but it is not the case with the usual expression for the
magnetic flux with the 3-vectors ($(e/\hbar )\int \mathbf{B}\cdot d\mathbf{S}
$); \emph{they cannot be equal.} The same objection refers to all other
relations with the 3-vectors in [1]. Hence, in this geometric approach,
using (\ref{cf}) and (\ref{FE}), the manifestly covariant expression for the
AB phase difference becomes%
\begin{equation}
\delta \alpha _{EB}=(-e/2\hbar )\int [(v_{\mu }E_{\nu }(x)-v_{\nu }E_{\mu
}(x))+\varepsilon _{\mu \nu \alpha \beta }v^{\alpha }B^{\beta }(x)]dx^{\mu
}\wedge dx^{\nu }.  \label{ps}
\end{equation}

In Sec. 3 in [1] it is investigated \textquotedblleft the usual magnetic AB
set-up of an infinite solenoid but with a time dependent magnetic field and
vector potential, i.e., $\mathbf{B}(t)$ and $\mathbf{A}(t)$%
.\textquotedblright\ As noted in [1] for that situation the scalar potential
is still zero, $\phi =0$. At first, it is worth mentioning that, as
explained in Sec. 3 in [3], in a correct covariant formulation there is no
static case. The 1-form $A$ ($A=A_{\mu }dx^{\mu }$) and the Faraday 2-form $%
F $ are both, always function of the position four-vector $x$; $A(x)$ and $%
F(x) $. If, for example, the usual 3-vector fields $\mathbf{A(\mathbf{r})}$,
$\mathbf{B(r)}$ do not explicitly depend on the coordinate time $t$ in one
frame, then the LT will mix the time and space coordinates; they cannot
transform the spatial coordinates from one frame only to spatial coordinates
in a relatively moving inertial frame of reference. What is static case for
one inertial observer is not more static case for relatively moving inertial
observer, but a time dependent case. Furthermore, if an observer uses the
\textquotedblleft radio\textquotedblright\ synchronization and not
Einstein's synchronization, then the space and time are not separated and
the usual 3-vector $\mathbf{\mathbf{r}}$ is meaningless. As can be seen from
Eq. (13) in [3] the components of the position 4-vector $x$ in the commonly
used coordinate basis with Einstein's synchronization and that one with the
\textquotedblleft radio\textquotedblright\ synchronization are connected as
\begin{equation}
x_{r}^{0}=x^{0}-x^{1}-x^{2}-x^{3},\quad x_{r}^{i}=x^{i},  \label{ptr}
\end{equation}%
and the same relation holds, e.g., for $(A_{r}^{0},A_{r}^{i})$, or $%
(E_{r}^{0},E_{r}^{i})$.

This consideration suggests that the results from Sec. 3 in [1] for the time
dependent, infinite solenoid, have to be reexamined using the correct
covariant formulation (\ref{ps}). We shall only discuss the AB phase
difference determined by Eqs. (8) and (9) in [1]. It is calculated using Eq.
(7) from [1]. This will be compared with (\ref{ps}). As already mentioned
above, in the 4D spacetime, Eq. (7) from [1] is not mathematically correct
and the same holds for Eqs. (8) and (9) from [1], which deal with the
3-vectors. The part of the AB phase difference with $B_{\mu }$\ from (\ref%
{ps}) is $\delta \alpha _{B}=(-e/2\hbar )\int \varepsilon _{\mu \nu \alpha
\beta }v^{\alpha }B^{\beta }(x)dx^{\mu }\wedge dx^{\nu }$ and it replaces
Eq. (8) from [1]. Only in the \textquotedblleft f\textquotedblright -frame
that part becomes $\delta \alpha _{B}=(-e/2\hbar )\int \varepsilon
_{0ijk}v^{0}B^{k}(x)dx^{i}\wedge dx^{j}$ and, as can be seen by the use of $%
B_{1}=B_{x}$, etc. that mathematically correct expression corresponds to Eq.
(8) from [1], i.e., to $\delta \alpha _{B}=(e/\hbar )\int \mathbf{B}\cdot d%
\mathbf{S}$. The essential difference is that all quantities in this
covariantly defined $\delta \alpha _{B}$ are properly defined in the 4D
spacetime and they correctly transform under the LT, like (\ref{el}), which
is not the case with the 3D quantities from Eq. (8) in [1].

The part of the AB phase difference with $E_{\mu }$\ from (\ref{ps}) is%
\begin{equation}
\delta \alpha _{E}=(-e/2\hbar )\int (v_{\mu }E_{\nu }(x)-v_{\nu }E_{\mu
}(x))dx^{\mu }\wedge dx^{\nu }  \label{de}
\end{equation}%
and, the same as for $\delta \alpha _{B}$, it is the same quantity for all
relatively moving inertial observers and for all bases used by them. Only in
the \textquotedblleft f\textquotedblright -frame $\delta \alpha _{E}$\ from (%
\ref{de}) becomes
\begin{equation}
\delta \alpha _{E}=(e/\hbar )\int v_{0}E_{i}(x)dx^{i}\wedge dx^{0}
\label{dae}
\end{equation}%
and it can be compared with Eq. (9) from [1]. For that comparison $F_{\mu
\nu }$ is written in terms of $A_{\mu }$\ as $F_{\mu \nu }=\partial _{\mu
}A_{\nu }-\partial _{\nu }A_{\mu }$. In that expression it is considered
that $A_{\mu }$\ are the primary quantities whereas $F_{\mu \nu }$ are
derived from them. But, as clearly shown in [15], the $F$ field is \emph{the
primary quantity} for the whole electromagnetism and not the four potential,
which is gauge dependent. However, here, for the comparison with [1], we use
the above relation with $A_{\mu }$. Then, (\ref{fc}) is used to get $E_{\mu
} $\ in terms of $A_{\mu }$, $E_{\mu }=F_{\alpha \mu }v^{\alpha }=(\partial
_{\alpha }A_{\mu }-\partial _{\mu }A_{\alpha })v^{\alpha }$. In the
\textquotedblleft f\textquotedblright -frame $E_{0}=0$\ and $E_{i}=(\partial
_{0}A_{i}-\partial _{i}A_{0})v^{0}$, what corresponds to the components of
the usual three-vector $\mathbf{E}$, e.g., $E_{1}$\ corresponds to $%
E_{x}=-\partial A_{x}/\partial t-\partial _{x}\phi $; remember that if $%
A_{\mu }$ is written in the usual notation it is $A_{\mu }=(\phi
,-A_{x},-A_{y},-A_{z})$ and in the \textquotedblleft f\textquotedblright
-frame $v^{\mu }=(1,0,0,0)$. Hence, in the \textquotedblleft
f\textquotedblright -frame, $\delta \alpha _{E}=(e/\hbar )\int (\partial
_{0}A_{i}-\partial _{i}A_{0})dx^{i}\wedge dx^{0}$, which for $A_{0}=0$
becomes $=(e/\hbar )\int \partial _{0}A_{i}dx^{i}\wedge dx^{0}$ and, by the
procedure from [1], it corresponds to Eq. (9) in [1], i.e., to $\delta
\alpha _{E}=(-e/\hbar )\int \mathbf{B}\cdot d\mathbf{S}=-\delta \alpha _{B}$%
. Thus, \emph{only} in the \textquotedblleft f\textquotedblright -frame and
for $A_{0}=0$ \textquotedblleft the two parts cancel
exactly.\textquotedblright\ Observe that the condition $A_{0}=0$ is not a
Lorentz covariant condition; in a relatively moving inertial frame $A_{0}$\
will be $\neq 0$. Furthermore, as seen from (\ref{ptr}), in the basis with
the \textquotedblleft radio\textquotedblright\ synchronization the temporal
and spatial components of $A_{\mu }$\ cannot be separated, which means that
in the 4D spacetime the condition $A_{0}=0$ has not a well-defined meaning.
The similar objections hold for the whole discussion presented in
[26].\bigskip \bigskip

\noindent \textbf{3. The Aharonov-Bohm effect in terms of fields}\bigskip

It is really surprising that both in all numerous theoretical discussions,
e.g., [1, 2, 27, 28, 26], in the experiments with microscopic solenoids [29]
and also in the recent experiment with macroscopic solenoid [30], it is
never noticed that in the rest frame of the solenoid there are \emph{always}
external \emph{static} electric fields for \emph{stationary,} \emph{resistive%
} conductor carrying \emph{constant current}. In an ohmic conductor there
are quasistatic surface charges, which generate not only the electric field
inside the wire driving the current, \emph{but also a time independent
electric field outside it}. That electric field is proportional to the
current, see, e.g., Sec. 4 in [12] and references therein. As mentioned in
[12] the existence of such quasistatic surface charges was first pointed out
by Kirchhoff, Refs. [18-20] in [12]. There are no analytic solutions for
these surface charges and the external electric fields for the case of
finite solenoids; for an infinite solenoid see [31]. The distribution of the
surface charges and the magnitude of the induced electric fields depend not
only on the geometry of the circuit but even of its surroundings. \emph{%
These external electric fields from steady currents} \emph{are firmly
experimentally confirmed}, see, e.g., [12], \emph{and they are well-known in
electrical engineering.} In [12], two other contributions to the external
electric field are discussed, but, as explained in Sec. 10 in [3], they are
of no concern here. It is worth mentioning that the expression from Sec. 4
in [12] is for a cylindrical wire of length $l$ carrying a constant current $%
I$ and that wire is a part of a square circuit. That expression is not
appropriate for a finite solenoid with steady current. In [31] an infinite
solenoid with steady current is considered and it is appropriate for the
case considered in [1]. There, in [31], a uniform cylindrical resistive
sheet of the radius $a$\ with a \textquotedblleft line\textquotedblright\
battery with terminals at potentials $\pm V_{0}/2$\ driving current \emph{%
azimuthally} in it is considered. In Sec. IV, [30], it is presented (i) the
magnitude of the electric field outside the solenoid, Eq. (11),
\begin{equation}
E=(V_{0}/\pi )(a/r\rho ),  \label{eis}
\end{equation}%
where $r$ and $\rho $ are the polar radii measured from the center (axis)
and from the baterry (respectively), and (ii) the electric lines of force,
Fig. 3. It is visible from Fig. 3. in [31] that the electric field has
radial and poloidal components, where the latter ones follow the direction
of the current just outside the solenoid in the same way as the magnetic
vector potential.

In the recent experiment [30] the absence of electromagnetic forces outside
the solenoid\ that are predicted by Boyer's force picture [32] has been
experimentally investigated by means of a time-of-flight experiment for a
macroscopic solenoid. It is looked for a time delay for electrons passing on
opposite sides of the solenoid. As discussed above in the generally accepted
theory the electron wave packets are influenced by nonzero vector potential,
i.e., by the quantum action of the magnetic flux even when electrons pass
through the field-free regions of space. On the other hand in Boyer's
semiclassical theory [32] there is a back-action force of the solenoid on
the electron, which gives rise to a time delay and to a phase shift that
exactly matches the AB-phase shift. It is shown in [30] that there is no
time delay and it is concluded\ that there are no fields predicted by
Boyer's force picture [32]. In his comment on the results obtained in [30]
Boyer [33] stated: \textquotedblleft the Aharonov-Bohm phase shift has never
been observed for such a macroscopic solenoid, .. .\textquotedblright\ In
[33], it is also argued that if the solenoid resistance is large, as in
[30], then the back forces will be small and there is no time lag, but for
the microscopic solenoids it is the opposite case. It has to be pointed out
that neither the authors of [30] nor Boyer [32, 33] knew anything about the
electric fields caused by the quasistatic surface charges that exist outside
the resistive conductors carrying \emph{constant} currents. This means that
it is not true that the paper [30] shows \emph{experimentally} that forces
cannot be responsible for the magnetic AB phase shift.\ The electric forces
caused by the mentioned quasistatic surface charges have nothing to do with
Boyer's force picture, [32, 33]. Thus, the main result from [30] about the
absence of the time delay does not imply that the electrons travel in a
field-free region. Obviously, the electric fields from quasistatic surface
charges have to be taken into account for the explanation of the AB phase
difference in the magnetic AB effect as well, i.e., in the usual magnetic AB
set-up of an infinite solenoid, which is considered in [1] and also in the
case of finite macroscopic [30] and microscopic [29] solenoids. From the
viewpoint presented here the AB phase difference in the magnetic AB effect
\emph{is not} due to the vector potential, i.e., according to Eq. (2) from
[1] due to the quantum action of the magnetic flux, but \emph{it is due to
the mentioned external electric field from stationary solenoids with steady
currents.} In that case, contrary to the generally accepted opinion, the
electron does not travel in the field-free region, but the electron wave
packets are \emph{locally }influenced by the electric field. A similar
expression as (\ref{dae}) is obtained in [34], Eq. (28), but their procedure
is not relativistically correct and the 3D electric field that enters into
their Eq. (28) is proportional to the square of the current.

In order to clarify the situation from the experimental viewpoint we
consider that some new experiments are required: the measurement in a \emph{%
single} experiment of the AB phase shift and the time delay, as suggested in
[33], and the measurement of the mentioned external electric fields \emph{%
separately} from AB-studies.

The above consideration implies that in the expression for $\delta \alpha
_{EB}$\ (\ref{ps}) there is no need to take into account the magnetic part $%
\delta \alpha _{B}$, i.e., the non-local effect of the magnetic field. In
the 4D spacetime only the local effects are important and physically
justified. This means that from the viewpoint of this approach with 4D
geometric quantities \emph{the AB phase difference is even for the magnetic
AB effect exclusively determined by the covariant expression} $\delta \alpha
_{E}$\ from (\ref{de}), i.e., by the local influence of the 4D electric
field. If the rest frame of the solenoid, the laboratory frame, is taken to
be the \textquotedblleft f\textquotedblright -frame then $\delta \alpha _{E}$%
\ is given by Eq. (\ref{dae}). In our opinion the magnetic part $\delta
\alpha _{B}$ of $\delta \alpha _{EB}$\ (\ref{ps}) could be taken into
account only in the case that the solenoid's magnetic field is not entirely
restricted to the coil's interior but exists in the coil's exterior as well,
i.e., along the electron's trajectory. The same conclusion that only the
local effect of the 4D electric field, i.e., $\delta \alpha _{E}$\ (\ref{de}%
) ((\ref{dae})) is important and physically meaningful holds in the same
measure for the time dependent set-up that is considered in Sec. 3 in [1].
Thus, in that case there is no cancellation of the non-local effect of the
magnetic field, $\delta \alpha _{B}$, with the local effect of the electric
field, $\delta \alpha _{E}$ (\ref{de}) ((\ref{dae})), because, as explained
above, only the electric field from the solenoid with current exists in the
region outside the solenoid and consequently it can locally influence the
electron travelling through that region. It is interesting that, as can be
seen from Sec. 4 in [35], if the current in the solenoid varies linearly
with time then it creates a time independent external electric field, see
Eq. (8) and Fig. 1 in [35]. Hence, for the solenoid with such a
time-dependent current there will be no time-dependent AB phase shift
although only $\delta \alpha _{E}$ (\ref{de}) ((\ref{dae})), the electric
part of $\delta \alpha _{EB}$\ (\ref{ps}), is considered to be physically
correct and justified.

Note that in this approach with the 4D geometric quantities the 3D
quantities from the usual approaches, e.g., from [1, 26, 12, 34, 35], etc.
have to be interpreted in a different way. Thus, for example, the \emph{%
components} of the electric field 3-vector in [1] have to be understood as
the \emph{spatial} \emph{components }in the standard basis of the 4D
electric field; the rest frame of the solenoid is taken to be the
\textquotedblleft f\textquotedblright -frame and therefore the temporal
component $E_{0}=0$ (also $B_{0}=0$). Also, in this geometric approach the
components $K_{\mu }$\ of the Lorentz force are given by (\ref{km}). As
discussed above, in the case considered in [1] only the electric part of $K$
from (\ref{km}) is physically important.\bigskip \bigskip

\noindent \textbf{4. Conclusions\bigskip }

As seen from the preceding discussion the correct covariant formulation of
the AB phase shift (\ref{ps})\ deals with the 4D geometric quantities that
properly transform under the mathematically correct LT (\ref{el}). In the 4D
spacetime Eq. (\ref{ps}) replaces Eqs. (8) and (9) from [1], which deal with
the 3D quantities that transform under the UT (\ref{ee}). Both, the 3D
quantities and their UT (\ref{ee}) are ill-defined in the 4D spacetime. As
\emph{proved} in [4-9], contrary to the generally accepted opinion, the UT (%
\ref{ee}) are not the mathematically correct LT. The main result that is
obtained in this paper is that \emph{even for the magnetic AB effect }(a
stationary solenoid with \emph{resistive} conductor carrying either steady
current or the current that varies linearly with time) \emph{the AB phase
difference is exclusively determined by the covariant expression} $\delta
\alpha _{E}$\ from (\ref{de}), i.e., by the \emph{local} influence of the 4D
electric field. Thus, here, it is shown that in the 4D spacetime only the
electric part of $\delta \alpha _{EB}$\ (\ref{ps}), i.e., $\delta \alpha
_{E} $\ (\ref{de}) ((\ref{dae})) is physically correct and meaningful. The
reason for it is that there are \emph{static} electric fields outside a
\emph{stationary,} \emph{resistive} conductor carrying \emph{steady current}%
, which means that it is not true that, e.g., in experiments [29, 30], the
electron travels in the field-free region of space. The existence of the
mentioned electric fields is firmly experimentally confirmed; for some
experiments see, e.g., [12] and references therein. All this together shows
that the magnetic AB phase shift considered in [1] is not a topological
phase shift.

In Sec. 7.1 in [3] it is shown that the external static electric fields, the
\textquotedblleft relativistic\textquotedblright\ second-order electric
fields, would need to exist not only for resistive conductors with steady
currents but even for superconducting solenoids with steady currents. In
Sec. 7.2 in [3] different experiments for the detection of the second-order
electric fields outside a stationary superconductor with steady current are
discussed. Furthermore, what is very important for the explanation of the AB
effect, in Sec. 8 in [3] such second-order electric fields are predicted to
exist outside a \emph{stationary} permanent magnet as well. As discussed in
Sec. 10 in [3], these results could explain the experimentally observed
fringe shift for the magnetic AB effect even in Tonomura's experiments [14]
\emph{in terms of previously overlooked electric forces and not,} \emph{as
generally accepted,\ in terms of nonzero vector potentials.}

Similarly, the qualitative theoretical explanations of the quantum phase
shifts in terms of the classical forces as the 4D vectors in the
Aharonov-Casher and the R\"{o}ntgen effects are presented in [7, 36].
Furthermore, in [37], the dipole moments are quantized and it is shown that
the expectation value for the quantum force 4D vector is not zero in the
case of the Aharonov-Casher and the R\"{o}ntgen effects and in the neutron
interferometry. Hence, in these experiments too the phase shifts are not due
to force-free interaction of the dipole, i.e., they also are not the
topological phase shifts.

The covariant AB effect in terms of $F$ and not in terms of a vector
potential is also investigated in [38].

It is interesting to note that recently another local explanation of the AB
effect is proposed in [39]. There, it is argued that if the solenoid in the
AB effect is treated in the framework of quantum theory then the effect can
be explained by the local action of the field of the electron on the
solenoid. In some respects there is a similarity between Boyer's calculation
[32, 33] and Vaidman's determination [39] of the AB phase shift. Boyer in
[32, 33] calculates the force exerted by the electron on a solenoid
(represented by a line of magnetic dipoles) and then relies on Newton's
third law to obtain a back-action force of the solenoid on the electron. The
same Boyer's force approach is investigated in [30] but a solenoid is
considered as a stack of current loops. However, Newton's third law is
violated for the electromagnetic interaction and to overcome this difficulty
a hidden momentum is often introduced, particularly in the case with current
loops, see, e.g. [30, 40] and references therein. But, as shown, e.g., in
[3] and [22], if an independent physical reality is attributed to the 4D
geometric quantities and not, as usual, to the 3D quantities, then there is
no need for the introduction of some \textquotedblleft
hidden\textquotedblright\ 3D quantities and there are no electromagnetic
paradoxes. Vaidman, in [39], see Fig. 4 in [39], considers that the electron
produces change in the magnetic flux of the solenoid, which causes an
electromotive force on charged solenoids (in his example). This leads to the
change in their velocity and to the shift of the wave packet of the
cylinders and finally to the correct expression for the AB phase, Eq. (5) in
[39] (arXiv: 1301.6153). \emph{Observe that this phase shift is for the
source (solenoids) and not for the passing electron.} Then, Vaidman states:
\textquotedblleft Since in quantum mechanics the wave function is for all
parts of the system together, the change of the wave function of the source
leads to observable effect in the interference experiment with the
electron.\textquotedblright\ (See Eqs. (8) and (9) in the first paper in
[39] for the change in the total wave function of the electron and the
solenoid.) It is worth noting that in Boyer's picture [32, 33] it is
impossible to detect the predicted force on the solenoid since it requires
the detection of the force of a single electron on a macroscopic object. For
the same reason, in Vaidman's picture [39], it is impossible to detect the
mentioned electromotive force and the change in the angular velocity of the
solenoids. Thus, both Boyer's force and Vaidman's electromotive force cannot
be experimentally verified, which means that neither of these approaches
have some physical, experimental, foundation.

On the other hand, the theory presented here is based on the existence of
the static \emph{electric fields} outside a \emph{stationary,} \emph{%
resistive} conductor carrying \emph{steady current}, and these fields are
already firmly experimentally verified.\bigskip

\noindent \textbf{Acknowledgments\bigskip }

It is a pleasure to acknowledge to Larry Horwitz and Martin Land for
inviting me to the IARD conferences and for their continuos support of my
work. I am also grateful to Zbigniew Oziewicz for numerous and very useful
discussions during years and to him and to Alex Gersten for the continuos
support of my work.\bigskip

\noindent \textbf{References\bigskip }

\noindent \lbrack 1] D. Singleton and E. C. Vagenas, Phys. Lett. B \textbf{%
723}, 241 (2013).

\noindent \lbrack 2] Y. Aharonov and D. Bohm, Phys. Rev. \textbf{115}, 485
(1959).

\noindent \lbrack 3] T. Ivezi\'{c}, J. Phys.: Conf. Ser. \textbf{437},
012014 (2013).

\noindent \lbrack 4] T. Ivezi\'{c}, Found. Phys. \textbf{33}, 1339 (2003).

\noindent \lbrack 5] T. Ivezi\'{c}, Found. Phys. Lett. \textbf{18}, 301
(2005).

\noindent \lbrack 6] T. Ivezi\'{c}, Found. Phys. \textbf{35}, 1585 (2005).

\noindent \lbrack 7] T. Ivezi\'{c}, Phys. Rev. Lett. \textbf{98},108901
(2007).

\noindent \lbrack 8] T. Ivezi\'{c}, arXiv: 0809.5277

\noindent \lbrack 9] T. Ivezi\'{c}, Phys. Scr. \textbf{82}, 055007 (2010).

\noindent \lbrack 10] J. D. Jackson, \textit{Classical Electrodynamics} 3rd
ed. (Wiley, New York, 1998).

\noindent \lbrack 11] H. Minkowski, Nachr. Ges. Wiss. G\"{o}ttingen, 53
(1908);

Reprinted in: Math. Ann. \textbf{68,} 472 (1910);

English translation in: M. N. Saha and S. N. Bose \textit{The Principle }

\textit{of Relativity: Original Papers by A. Einstein and H. Minkowski}

(Calcutta University Press, Calcutta, 1920).

\noindent \lbrack 12] A. K. T. Assis, W. A. Rodrigues Jr. and A. J. Mania,

Found. Phys. \textbf{29}, 729 (1999).

\noindent \lbrack 13] T. Ivezi\'{c}, Phys. Scr.\textit{\ }\textbf{81,}
025001 (2010).

\noindent \lbrack 14] A. N. Tonomura, T. Osakabe, T. Matsuda, T. Kawasaki,
J. Endo,

S. Yano, and H. Yamada, Phys. Rev. Lett. \textbf{56}, 792 (1986).

\noindent \lbrack 15] T. Ivezi\'{c}, Found. Phys. Lett. \textbf{18,} 401
(2005).

\noindent \lbrack 16] T. Ivezi\'{c}, arXiv: 1101.3292

\noindent \lbrack 17] A. Einstein, Annalen der Physik \textbf{17,} 891
(1905); Translated by W. Perrett

and G. B. Jeffery in: \textit{The Principle of Relativity} (Dover, New York,
1952).

\noindent \lbrack 18] T. Ivezi\'{c}, Found. Phys. \textbf{31,} 1139 (2001).

\noindent \lbrack 19] D. A. T. Vanzella, Phys. Rev. Lett. \textbf{110},
089401 (2013).

\noindent \lbrack 20] H. N. N\'{u}\~{n}ez Y\'{e}pez, A. L. Salas Brito and
C. A. Vargas, Revista Mexicana

de F\'{\i}sica \textbf{34}, 636 (1988); S. Esposito, Found. Phys. \textbf{28}%
, 231 (1998);

J. Anandan, Phys. Rev. Lett. \textbf{85}, 1354 (2000).

\noindent \lbrack 21] R. M. Wald, \textit{General Relativity} (The
University of Chicago Press,

Chicago, 1984); M. Ludvigsen, \textit{General\ Relativity}, \textit{A
Geometric Approach}

(Cambridge University Press, Cambridge, 1999);

S. Sonego and M. A. J. Abramowicz, J. Math. Phys. \textbf{39}, 3158 (1998);

D. A. T. Vanzella, G. E. A. Matsas, H. W. Crater, Am. J. Phys. \textbf{64},

1075 (1996); Z. Oziewicz, J. Phys.: Conf. Ser.\textit{\ }\textbf{330},
012012 (2011);

F. W. Hehl and Yu. N. Obukhov, \textit{Foundations of Classical }

\textit{Electrodynamics: Charge, flux, and metric} (Birkh\"{a}user, Boston,
2003);

R. D. Blandford and K. S. Thorne, \textit{Applications of classical physics}

(California Institute of Technology, 2002-2003),

[http://www.pma.caltech.edu/Courses/ph136/yr2004/].

\noindent \lbrack 22] T. Ivezi\'{c}, arXiv: 1212.4684

\noindent \lbrack 23] T. Ivezi\'{c}, Found. Phys. Lett. \textbf{15}, 27
(2002);

T. Ivezi\'{c}, arXiv: physics/0103026;

T. Ivezi\'{c}, arXiv: physics/0101091.

\noindent \lbrack 24] T. Ivezi\'{c}, Found. Phys. \textbf{36}, 1511 (2006);

T. Ivezi\'{c}, Fizika A \textbf{16}, 207 (2007).

\noindent \lbrack 25] T. Ivezi\'{c}, Found. Phys. \textbf{37}, 747 (2007).

\noindent \lbrack 26] J. Macdougall and D. Singleton, J.Math.Phys. \textbf{55%
}, 042101 (2014).

\noindent \lbrack 27] M. Peshkin and A. Tonomura, \textit{The Aharonov-Bohm
Effect}

(Springer, New York, 1989).

\noindent \lbrack 28] Y. Aharonov and D. Rohrlich, \textit{Quantum Paradoxes}%
, \textit{Quantum Theory for }

\textit{the Perplexed} (Wiley-VCH, Weinheim, 2005).

\noindent \lbrack 29] G. M\"{o}llenstedt and W. Bayh, Naturwissenschaften
\textbf{49}, 81 (1962).

\noindent \lbrack 30] A. Caprez, B. Barwick and H. Batelaan, Phys. Rev.
Lett. \textbf{99},

210401 (2007).

\noindent \lbrack 31] M. A. Heald, Am. J. Phys. \textbf{52}, 522 (1984).

\noindent \lbrack 32] T. H. Boyer, Found. Phys. \textbf{32}, 41 (2002).

\noindent \lbrack 33] T. H. Boyer, Found. Phys. \textbf{38}, 498 (2008).

\noindent \lbrack 34] H. Torres Silva and A. K. T. Assis, Revista Facultad
de Ingenieria

U. T. A. (Chile) \textbf{9}, 29 (2001).

\noindent \lbrack 35] M. A. Miller, Sov. Phys. Usp. \textbf{27}, 69 (1984).

\noindent \lbrack 36] T. Ivezi\'{c}, Phys. Rev. Lett. \textbf{98, }158901
(2007).

\noindent \lbrack 37] T. Ivezi\'{c}, arXiv: hep-th/0705.0744.

\noindent \lbrack 38] J. Anandan, Int. J. Theor. Phys. \textbf{19}, 537
(1980).

\noindent \lbrack 39] L. Vaidman, Phys. Rev. A \textbf{86, }040101 (R)
(2012);

L. Vaidman, \textit{Quantum Theory: A Two-Time Success Story - Yakir Aharonov%
}

\textit{Festschrift}, D. C. Struppa and J. M. Tollaksen, eds., pp. 247-255

(Springer, Italia, 2014);

L. Vaidman, arXiv: 1301.6153.

\noindent \lbrack 40] S. McGregor, R. Hotovy, A. Caprez and H. Batelaan, New
J. Phys. \textbf{14},

093020 (2012).

\end{document}